\documentstyle[aps,prl,multicol,floats,epsf]{revtex}

\begin{document}
\draft
\wideabs{

\title{Magnetotransport properties of untwinned 
YBa$_{2}$Cu$_{3}$O$_{y}$ single crystals: \\
Novel 60-K-phase anomalies in the charge transport}

\author{Yoichi Ando\cite{corr} and Kouji Segawa}
\address{Central Research Institute of Electric Power Industry, 
Komae, Tokyo 201-8511, Japan}

\date{\today}
\maketitle

\begin{abstract}
We present the result of our accurate measurements of 
the  $a$- and $b$-axis resistivity ($\rho_a$ and $\rho_b$), 
magnetoconductivity $\Delta\sigma/\sigma$, 
Hall coefficient $R_H$, and the $a$-axis thermopower $S_a$ 
in untwinned YBa$_{2}$Cu$_{3}$O$_{y}$ single crystals in a 
wide range of doping ($6.45 \le y \le 7.0$).  
The systematics of our data reveals a number of 
novel 60-K-phase anomalies in the charge transport: 
(i) Temperature dependences of $\rho_a$ show anomalous 
overlap below $\sim$130 K for $6.65 \le y \le 6.80$,
(ii) Hall mobility $\mu_{H}$ shows an enhancement 
near $y \simeq 6.65$, which is reflected in an 
anomalous $y$ dependence of $\sigma_{xy}$, 
(iii) With decreasing temperature $R_H$ shows a marked drop 
upon approaching $T_c$ only in samples with $6.70 \le y \le 6.85$, 
(iv) Superconducting fluctuation magnetoconductivity is 
anomalously enhanced near $y \simeq 6.7$, and 
(v) $H_{c2}$ is anomalously reduced near $y \simeq 6.70$. 
We discuss that the fluctuating charge stripes might be 
responsible for these anomalies in the charge transport.
\end{abstract}

\pacs{Keywords: Transport Properties, 60-K phase, Superconducting 
Fluctuations, Coherence Length}
}
\narrowtext

\section{Introduction}

In YBa$_{2}$Cu$_{3}$O$_{y}$ (YBCO) compound, 
an increase in the oxygen content $y$ from 6 to 7 causes 
the hole doping into the CuO$_2$ planes (above $y \simeq 6.2$) 
and leads to superconductivity (above $y \simeq 6.4$).
The dependence of $T_c$ on $y$ is non-trivial 
and there is a well-known plateau with $T_c \simeq$ 60 K 
for $y$ of around 6.7 (``60-K plateau" or ``60-K phase").
Both an oxygen-ordering scenario \cite{Veal} and an 1/8-anomaly 
scenario \cite{Tallon} have been discussed for the 
origin of the 60-K plateau, but the case remains controversial.
To understand the true nature of the 60-K phase in YBCO, 
we have conducted systematic measurements of the transport 
properties across the 60-K phase, which reveal
clear electronic anomalies in the 60-K phase and thus indicate 
an electronic origin of the 60-K plateau. 
In this paper, we summarize the novel 60-K-phase anomalies 
we found in the charge transport in YBCO and discuss their 
possible origin in conjunction with the self-organization of 
the holes into stripes.

\begin{figure}[t!]
\epsfxsize=0.95\columnwidth
\centerline{\epsffile{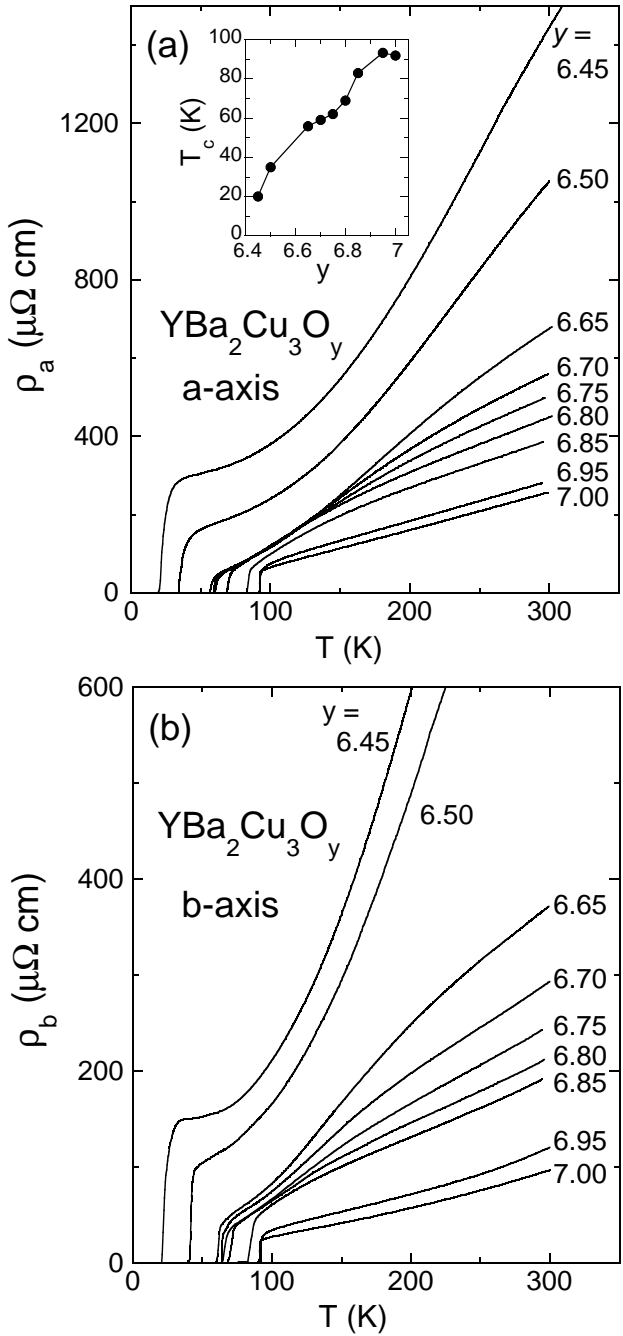}}
\vspace{0.2cm}
\caption{$T$ dependences of (a) $\rho_a$ and (b) $\rho_b$ for 
untwinned YBCO crystals in 0 T.  
Inset in (a): Phase diagram of zero-resistance $T_c$ vs. $y$.}
\label{fig1}
\end{figure}

\section{Experimental}

It should be noted that the oxygen arrangement in the Cu-O 
chain layers is essentially random, causing 
complications to the study of YBCO; for example, for a given 
$y$ the actual hole doping can differ 
depending on the arrangement of the O atoms, and 
the O atoms in the Cu-O chains can rather easily rearrange at 
room temperature, which causes the ``room-temperature (RT) 
annealing effect" \cite{Lavrov}.  
For this work, the crystals are always quenched at the 
end of the high-temperature annealing (which tunes the oxygen 
content) and detwinning is performed at temperatures below 
220$^{\circ}$C after the annealing.
The samples are left at room temperature for 
at least a week for the oxygen arrangement to equilibrate before 
the measurements; therefore, the oxygen atoms on the chain sites 
are expected to be locally ordered (because of the RT annealing) 
but macroscopically uniform (because of the quenching) \cite{Lavrov}.
(Note that a long-time annealing at relatively low temperature 
($\sim$100$^{\circ}$C) causes a macroscopic phase separation in 
heavily-underdoped samples \cite{Radaelli} and messes up the 
transport properties.)
Our procedure ensures very good reproducibility of the 
transport properties, as has been demonstrated in Ref. \cite{60K}.  
In particular, we have determined the absolute values 
of the resistivity and the Hall coefficient for a given $y$ with 
the accuracy of 5\%; this gives us confidence in discussing the 
systematics of the transport properties across the 60-K phase.
Details of the measurement techniques are described in Refs. 
\cite{60K,Abe,stripeMR}.

\section{Results and Discussions}

Let us start with the resistivity behavior.  
Figure 1(a) shows the temperature dependences of $\rho_a$ for a wide 
range of oxygen contents, $y = 6.45-7.0$.  Remember that 
the Cu-O chains run along the $b$-axis and thus $\rho_a$ is not 
complicated with the conductivity of the chains \cite{Gagnon}.
We emphasize that at least 3 samples are measured for each 
$y$ and the data are reproducible within 5\%; in fact, 
Fig. 1(a) is a summary of the measurements of more than 30 samples.
One can immediately notice in Fig. 1(a) that the $\rho_a(T)$ data for 
$y = 6.65-6.80$ show clear overlap below $\sim$130 K.
Note that in the underdoped YBCO the pseudogap opening 
can be inferred from a downward deviation from the 
high-temperature $T$-linear dependence \cite{Ito}, 
and thus the overlapping of $\rho_a(T)$ is observed 
in the pseudogap state. 
Figure 1(b) shows the corresponding evolution of $\rho_b(T)$ 
with $y$; $\rho_b$ is generally smaller than 
$\rho_a$ for the same $y$, which is believed to be due to 
the finite chain conductivity \cite{Gagnon}.
The $\rho_b(T)$ data do not show as clear overlap in the 
60-K phase as the $\rho_a(T)$ data. 

\begin{figure}[t!]
\epsfxsize=0.8\columnwidth
\centerline{\epsffile{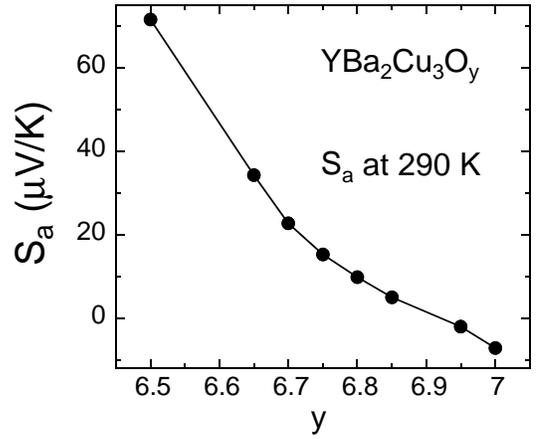}}
\vspace{0.2cm}
\caption{$y$ dependence of the $a$-axis thermopower at 290 K.}
\label{fig2}
\end{figure}

The overlap of the $\rho_a(T)$ data is very unusual.  
Unless the effective mass $m^*$ of the charge carrier in YBCO 
is anomalously changing with $y$ (which is very unlikely), 
the $y$-independence of $\rho_a$ can have only 
two possible origins: (i) both the carrier concentration $n$ 
and the scattering time $\tau$ remain unchanged with $y$, or 
(ii) a change in $n$ is compensated by a change in $\tau$.  
To clarify which of the two is actually the case, we measured 
the room-temperature thermopower $S(290{\rm K})$, which is 
generally believed to reflect the change in the hole concentration 
and thus may be used as a guide to estimate $n$ \cite{Tallon}. 
Figure 2 shows the $y$ dependence of the thermopower measured 
along the $a$-axis at 290 K, $S_a(290{\rm K})$, which is expected 
to be free from the contribution of the Cu-O chain transport.
The $S_a(290{\rm K})$ data show a continuous change across the 
60-K phase, which strongly suggests that $n$ is continuously 
changing with $y$ in our samples.  Therefore, among the two 
possibilities listed above, it is more likely that a change in 
$n$ is somehow compensated by a change in $\tau$ in the 60-K phase.

\begin{figure}[t!]
\epsfxsize=0.8\columnwidth
\centerline{\epsffile{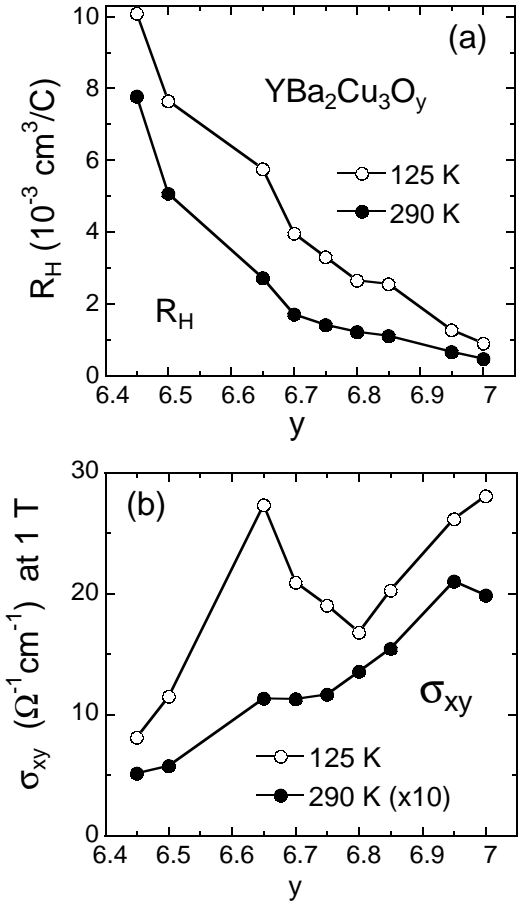}}
\vspace{0.2cm}
\caption{$y$ dependences of (a) raw $R_H$ and (b) Hall conductivity 
$\sigma_{xy}$ (calculated for $B$=1 T) at 125 and 290 K.}
\label{fig3}
\end{figure}

Examination of the Hall channel in the in-plane transport 
gives us a clue to the (phenomenological) origin of the 
anomaly in $\rho_a(T)$.  
In YBCO, it is expected that $\sigma_{xy}$ is governed by the 
properties of the planes (since the Cu-O chains contribute 
little to $\sigma_{xy}$ because of their one-dimensionality), 
while the Hall resistivity $\rho_{xy}$ [which is expressed 
as $\rho_{xy} \simeq \sigma_{xy}/(\sigma_{xx}\sigma_{yy})$ with 
$\sigma_{xx} \simeq 1/\rho_a$ and $\sigma_{yy} \simeq 1/\rho_b$] 
is affected by the properties of the chains through $\sigma_{yy}$.
Therefore, $\sigma_{xy}$ is a better indicator of the 
properties of the planes compared to $R_H$. 
Perhaps reflecting this situation, $\sigma_{xy}$ at 125 K shows 
a clearly anomalous $y$ dependence while such anomalies are 
smeared out in the $y$ dependence of $R_H$ [Figs. 3(a) and 3(b)].
The nature of this anomaly in the Hall channel is 
best understood by the plot of the Hall mobility in the planes,
$\mu_H = \sigma_{xy}/(B\sigma_{xx})$ (Fig. 4), 
which reflects $\tau /m^*$ and does not include $n$ in the 
Drude picture.  One can clearly see that $\mu_H$ is anomalously 
enhanced near $y$=6.65, particularly at 125 K 
[where the overlap of $\rho_a(T)$ is observed].
Therefore, it appears that the scattering time $\tau$ gets enhanced 
upon reducing $y$ from 6.80 to 6.65 and this change in $\tau$ 
cancels the change in $n$, causing the $y$-independent resistivity 
in the CuO$_2$ planes for $y = 6.65-6.80$.  
It should be noted that this anomalous enhancement of $\tau$ takes 
place only when the pseudogap opens, and thus the overlap 
of $\rho_a(T)$ is observed only below $\sim$130 K.

\begin{figure}[b!]
\epsfxsize=0.8\columnwidth
\centerline{\epsffile{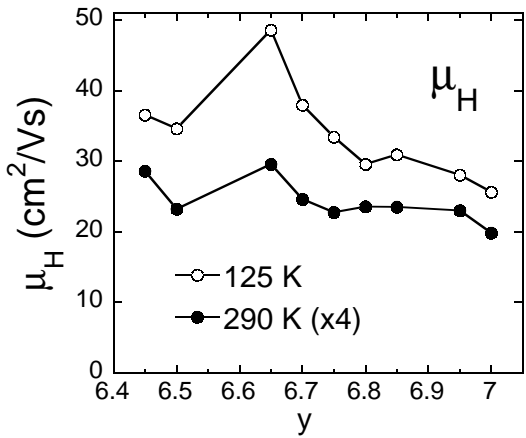}}
\vspace{0.2cm}
\caption{$y$ dependence of the Hall mobility at 125 and 290 K.}
\label{fig4}
\end{figure}

\begin{figure}[t!]
\epsfxsize=1\columnwidth
\centerline{\epsffile{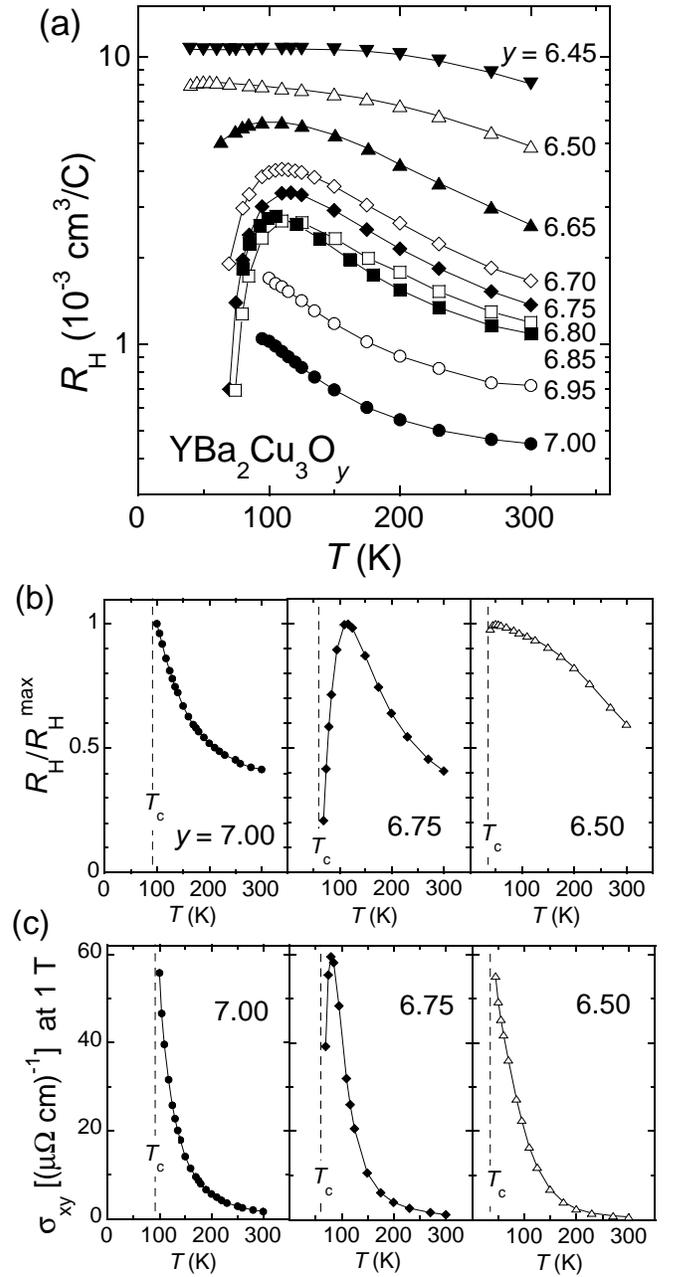}}
\vspace{0.2cm}
\caption{(a) $T$ dependences of $R_H$ for various $y$. 
(b), (c): $T$ dependences of $R_H$ and $\sigma_{xy}$ (calculated 
for $B$=1 T) for samples with $y=7.00$, 6.75, and 6.50; 
the $R_H$ data are normalized by the maximum values for each $y$ 
in (b).}
\label{fig5}
\end{figure}

The Hall effect reveals yet another aspect of the anomalies in 
the 60-K phase.
Figure 5(a) shows the temperature dependences of $R_H$ (measured with 
the current along the $a$-axis) for a wide range of $y$; the data 
points in Fig. 3(a) are extracted from this series of data.
One may notice in Fig. 5(a) that only the data for $y = 6.65-6.85$ 
show noticeable drop in $R_H(T)$ with lowering temperature upon 
approaching $T_c$; this situation becomes 
obvious in Fig. 5(b), where the $R_H(T)$ data for $y=7.0$, 6.75, 
and 6.50 are compared, with clearly indicated $T_c$.
It is apparent that only the 60-K-phase sample shows a marked drop 
in $R_H(T)$ from well above $T_c$ and the data seem to be heading 
towards zero.  To demonstrate that this anomaly is not extrinsically 
caused by the effect of the Cu-O chains on $R_H$, we show similar 
comparison of $\sigma_{xy}(T)$ for the three compositions [Fig. 5(c)], 
which again shows that the anomalous $T$ dependence is observable 
(though a bit weakened) only in the 60-K phase sample.
This remarkable drop in $R_H$ is somewhat reminiscent of the 
$R_H(T)$ behavior in Nd-doped La$_{2-x}$Sr$_{x}$CuO$_{4}$ \cite{Noda}, 
where it has been discussed that some peculiarity of the 
charge transport \cite{Noda,Emery,Prelovsek} imposed by the 
stripes \cite{Tranquada} is responsible for the drop in $R_H$.
Thus, the anomaly observed in the $R_H(T)$ behavior is suggestive 
of a role of the stripes in the 60-K phase of YBCO.

\begin{figure}[t!]
\epsfxsize=1\columnwidth
\centerline{\epsffile{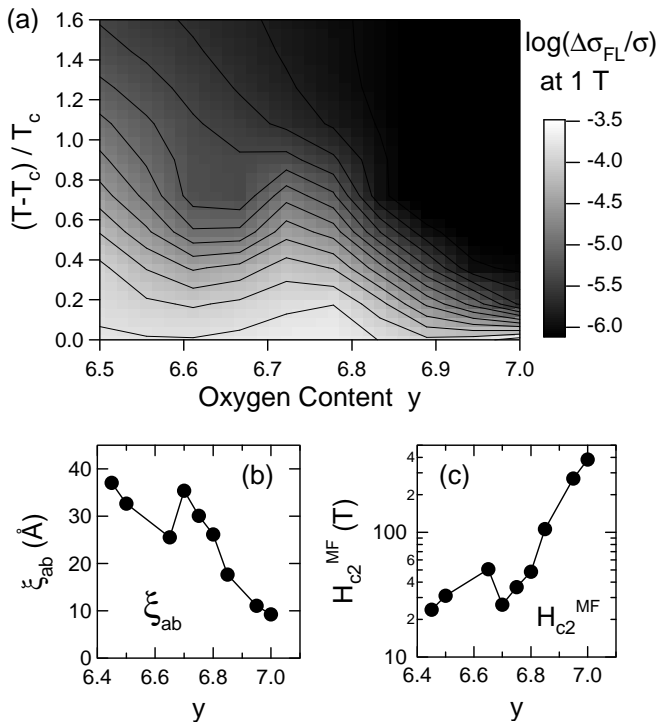}}
\vspace{0.2cm}
\caption{(a) Evolution of the fluctuation MC in the 
$(T-T_c)/T_c$ vs. $y$ plane. (b) $y$ dependence of 
$\xi_{ab}$ obtained from the ALO fits. (c) $y$ dependence of 
the mean-field upper critical field calculated from $\xi_{ab}$.}
\label{fig6}
\end{figure}

We also measured the magnetoresistance in $\rho_a$ in both 
the transverse and the longitudinal geometries, from which we 
extract orbital magnetoconductivity (MC).  The orbital MC at 
high temperatures, whose temperature dependence is very well fitted 
with $[bT^2+c]^{-2}$ \cite{Harris,MR}, is due to the normal-state 
contribution; the data show additional contribution at lower 
temperatures, which can be identified as the Aslamazov-Larkin orbital 
(ALO) component of the superconducting fluctuation MC \cite{Harris}.  
The full details of the data and their analysis will be 
published elsewhere \cite{MR}, and here we show only the fluctuation 
MC ($\Delta\sigma_{\rm FL}/\sigma$) obtained after the analysis.  
Figure 6(a) shows how the fluctuation MC evolves in the 
$(T-T_c)/T_c$ vs. $y$ plane.  
One can see that there is a general growth of 
$\Delta\sigma_{\rm FL}/\sigma$ with decreasing $y$, but on top of 
this trend there is a marked enhancement of 
$\Delta\sigma_{\rm FL}/\sigma$ near $y \simeq 6.7$.  
An enhancement of $\Delta\sigma_{\rm FL}/\sigma$ means that 
the characteristic magnetic-field scale to suppress the 
superconducting fluctuations is smaller, which suggests that 
$H_{c2}$ is reduced in the 60-K phase. 
In fact, when we extract the in-plane coherence length $\xi_{ab}$
by fitting the $\Delta\sigma_{\rm FL}/\sigma$ data to the 
ALO formula, obtained $\xi_{ab}$ [Fig. 6(b)] shows an 
anomalous enhancement in its $y$ dependence near $y \simeq 6.7$;
this gives rise 
to a reduction of the mean-field upper critical field 
$H_{c2}^{\rm MF}$ [$\equiv \Phi_0/(2\pi\xi_{ab}^2)$] as 
shown in Fig. 6(c).

\section{Stripes?}

All the above results indicate that the {\it electronic} state 
in the CuO$_2$ planes near $y \simeq 6.7$ is somewhat anomalous 
compared to other composition and that the 60-K-phase anomalies 
are definitely not simply due to oxygen ordering. 
The origin of the electronic anomaly in the 60-K phase is 
not clear at this stage, but an intriguing possibility is the 
charge stripes, because the drop of $R_H$ with decreasing 
temperature observed in the 60-K-phase samples is most easily 
associated with the stripe physics \cite{Noda,Emery,Prelovsek}.
Below we propose a highly speculative scenario, which just 
describes one possibility:
Suppose that the physics in the whole phase diagram is governed 
by the fluctuating stripes or ``electronic liquid crystals" 
(as is proposed by Kivelson {\it et al.} \cite{Kivelson}) and that 
the superconductivity is caused by the fluctuating stripes.
If there is a quantum critical point (QCP) associated with the 
stripes near $y \simeq 6.7$ \cite{Kivelson2}, 
the superconductivity may be weakened at such QCP (which leads to 
a reduction in $H_{c2}$ and a plateau in $T_c$). 
Since the fluctuations are always enhanced near QCP, the charge 
mobility may also be enhanced.  
Thus, though highly speculative, it is possible to 
qualitatively understand the observed anomalies in the 60-K phase 
in terms of the electronic liquid crystal picture.
In the above scenario, the drop in $R_H(T)$ should be due to 
the particle-hole symmetry \cite{Emery,Prelovsek} rather than the 
suppression of the transverse motion of charges \cite{Noda}. 

\section{Summary}

We have systematically measured the transport 
properties of untwinned YBCO single crystals in a wide range of 
doping and found that the behaviors of $\rho_a(T)$, 
$\mu_{H}(y)$, $R_H(T)$,  $\Delta\sigma_{\rm FL}(y)$, and 
$H_{c2}^{\rm MF}(y)$ all show novel anomalies in the 60-K 
phase.  These anomalies are clearly of electronic origin, and 
possibly related to the stripe physics in this compound.

\section{Acknowledgments}

We thank S. A. Kivelson and A. N. Lavrov for helpful discussions.

%
\medskip
\vfil


\vspace{-0.8cm}

\begin{references}

\vspace{-1.6cm}

\bibitem[*]{corr}
Corresponding author. Fax: +81-3-3480-3401. \\
{\it E-mail address:} ando@criepi.denken.or.jp (Y. Ando).

\vspace{0.4cm}

\bibitem{Veal}
B. W. Veal and A. P. Paulikas, Physica C 184 (1991) 321, 
and references therein.

\bibitem{Tallon}
J. L. Tallon, G. V. M. Williams, N. E. Flower, and C. Bernhard, 
Physica C 282-287 (1997) 236.

\bibitem{Lavrov}
A. N. Lavrov and L. P. Kozeeva, Physica C 253 (1995) 313, 
and references therein.

\bibitem{Radaelli}
P. G. Radaelli, C. U. Segre, D. G. Hinks, and J. D. Jorgensen,
Phys. Rev. B 45 (1992) 4923.

\bibitem{60K}
K. Segawa and Y. Ando, Phys. Rev. Lett. 86 (2001) 4907.

\bibitem{Abe}
Y. Abe, K. Segawa, and Y. Ando, Phys. Rev. B 60 (1999) R15055.

\bibitem{stripeMR}
Y. Ando, A. N. Lavrov, and K. Segawa, Phys. Rev. Lett. 83 (1999) 2813.

\bibitem{Gagnon}
R. Gagnon, C. Lupien, and L. Taillefer, Phys. Rev. B 50 (1994) 3458.

\bibitem{Ito}
T. Ito, K. Takenaka, and S. Uchida, Phys. Rev. Lett. 70 (1993) 
3995.

\bibitem{Noda}
T. Noda, H. Eisaki, and S. Uchida, Science 286 (1999) 265.

\bibitem{Emery}
V. J. Emery, E. Fradkin, S. A. Kivelson, and T. C. Lubensky,
Phys. Rev. Lett. 85 (2000) 2160.

\bibitem{Prelovsek}
P. Prelovsek, T. Tohyama, and S. Maekawa, cond-mat/0102418.

\bibitem{Tranquada}
J. M. Tranquada, B. J. Sternlieb, J. D. Axe, Y. Nakamura, and 
S. Uchida, Nature 375 (1995) 561.

\bibitem{Harris}
J. M. Harris {\it et al.}, Phys. Rev. Lett. 75 (1995) 1391.

\bibitem{MR}
Y. Ando and K. Segawa, cond-mat0108054.

\bibitem{Kivelson}
S. A. Kivelson, E. Fradkin, and V. J. Emery, Nature 393 (1998) 550.

\bibitem{Kivelson2}
S. A. Kivelson (private communication).

\end{references}
\end{document}